\documentclass[runningheads]{llncs}
\usepackage[T1]{fontenc}
\usepackage{latexsym}

\usepackage{tabulary,graphicx,amsmath,amsfonts,amssymb,setspace}
\usepackage{booktabs}
\usepackage{tablefootnote}
\usepackage{tabularx}
\usepackage{multirow}
\usepackage{float}
\usepackage{hyperref} 
\hypersetup{
hidelinks,
colorlinks=true,
linkcolor=blue,
citecolor=blue,
urlcolor = black
}
\usepackage[dvipsnames,table]{xcolor}

\usepackage{caption}

\usepackage{color}

\begin{document}
\title{Exploring Efficient-Tuned Learning Audio Representation Method from BriVL
}
%
%
%
%
%

\author{Sen Fang\inst{1} \and Yangjian Wu\inst{2} \and Bowen Gao\inst{1} \and Jingwen Cai\inst{1} \and Teik Toe Teoh\inst{3,\star}} \authorrunning{F. Sen et al.}

\institute{Victoria University \email{\{sen.fang, bowen.gao, jingwen.cai\}@live.vu.edu.au} \and Hainan University \email{yangjian.wu@hainanu.edu.cn} \and Nanyang Technological University \email{ttteoh@ntu.edu.sg}}

\maketitle              
\begin{abstract}
Recently, there has been an increase in the popularity of multimodal approaches in audio-related tasks, which involve using not only the audible modality but also textual or visual modalities in combination with sound. 
In this paper, we propose a robust audio representation learning method WavBriVL based on Bridging-Vision-and-Language (BriVL). 
It projects audio, image and text into a shared embedded space, so that multi-modal applications can be realized. 
We tested it on some downstream tasks and presented the images rearranged by our method and evaluated them qualitatively and quantitatively. The main purpose of this article is to:
(1) Explore new correlation representations between audio and images;
(2) Explore a new way to generate images using audio.
The experimental results show that this method can effectively do a match on the audio image.

\keywords{Audio-Visual  \and Multimodal Learning \and Generative Adversarial Network \and Speech Representation Learning.}
\end{abstract}

\section{Introduction}

Data volumes are crucial for training large-scale language models, and the use of vast corpora has been the standard practice \cite{kaplan2020scaling}. 
However, the limited availability of high-quality labeled data in both unimodal and multimodal directions has hindered the development of the field \cite{devlin2018bert}. 
To address this constraint, researchers have turned to zero-shot and few-shot learning approaches that rely on contrastive learning methods using textual descriptions.
While the use of additional modalities together with text is common, the combination of more than two modalities in the audio domain is still uncommon. 
Given that manual data annotation in supervised learning is prohibitively expensive, self-supervised learning holds significant value for training large models. 
To overcome resource constraints and expand the research field, we present a novel multimodal self-supervised model based on the cutting-edge work of \textbf{Bridging-Vision-and-Language} \cite{fei2022towards}.

On the basis of this work, we propose an audio-visual correspondence model that extracts training from the BriVL model.
BriVL is a text image correspondence model similar to OpenAI CLIP \cite{radford2021learning} and Google ALIGN \cite{jia2021scaling}, published in Nature Communications.
Like CLIP, BriVL can rearrange images based on how well they match text images to find the best match.
The principle of Our straightforward procedure is to freeze the BriVL visual encoder and part weight, run video on the visual stream of the model, and train a new model to predict BriVL embedding independently from the audio stream. 
The method we talked about above is very simple, which can secondary train, output images, and expand on more tasks. 
In addition, WavBriVL embeddings originate from BriVL, which means they align with text. 
This makes audio guided image repair \cite{zhao2022generating}, audio subtitles and cross mode text/audio to audio/image retrieval possible. 
We found WavBriVL is easy to extend to large video datasets. 
We systematically evaluated WavBriVL in a variety of audio tasks, including classification and retrieval, and compared it with other audio representation learning methods, as well as the SOTA results of each task. 
We showed how to apply WavBriVL to solve multiple multi-modal and zero start tasks. 

\vspace{-1.0em}
\section{Related Works}

Our work is motivated by recent advancements in the field of multimodal learning, particularly in the first half of 2022. BriVL has shown to outperform CLIP \cite{radford2021learning} in various benchmarks and Microsoft's new WavLM \cite{chen2022wavlm} has surpassed their previous Wav2Vec \cite{baevski2020wav2vec} in multiple aspects. We hypothesize that combining these two models could achieve better results than Wav2CLIP \cite{wu2022wav2clip}. 
Research on giving AI multi-modal perception and reasoning has been ongoing for many years, however, image generation for audio input is still an emerging field.
With the emergence of different generation models, such as Goodfellow introduced GAN in 2014, there has been a lot of excellent work in the field of image generation \cite{karras2017audio,cudeiro2019capture,yi2020audio,zhang20213d,song2022everybody,zhang2021facial,zhang2021flow,wu2021imitating,lahiri2021lipsync3d,richard2021meshtalk,thies2020neural}.
From single mode to multi-modal, from text guidance about 2015 years later to audio guidance 2020 years later, there are many impressive works \cite{qiu2018image,zhu2021deep}.
Naturally, there exist numerous antecedent endeavors and unacknowledged contributions. 
After this, there have been significant advancements in the field of image generation. 
Some of the most popular techniques include Stable Diffusion models \cite{Rombach_2022_CVPR} and ControlNet \cite{zhang2023adding},
a diverse range of applications has utilized these techniques, showcasing their potential in generating high-quality images. 

In the work of modal extension for CLIP, AudioCLIP and related works like Wav2CLIP\footnote{\url{https://github.com/descriptinc/lyrebird-wav2clip}} are similar but different from our work. Compared to Wav2CLIP, BriVL has less dependence on specific semantic relationships, resulting in more creativity. Compare with them, we adopted a simple and efficient dual-tower architecture that does not rely on time-consuming target detectors. Finally, BriVL incorporates a cross-modal comparative learning algorithm based on MoCo \cite{9157636}, which differs from the approach in CLIP.
Overall, while our work builds upon existing multi-modal learning research, and temporarily verify our idea on image generation area. 
We introduce novel techniques and a unique approach that contributes to the field of audio-guided image generation.
Furthermore, we conducted a comprehensive series of multimodal task tests that successfully demonstrated the efficacy of our proposed method.

\begin{figure}[htbp]
\includegraphics[width=\linewidth]{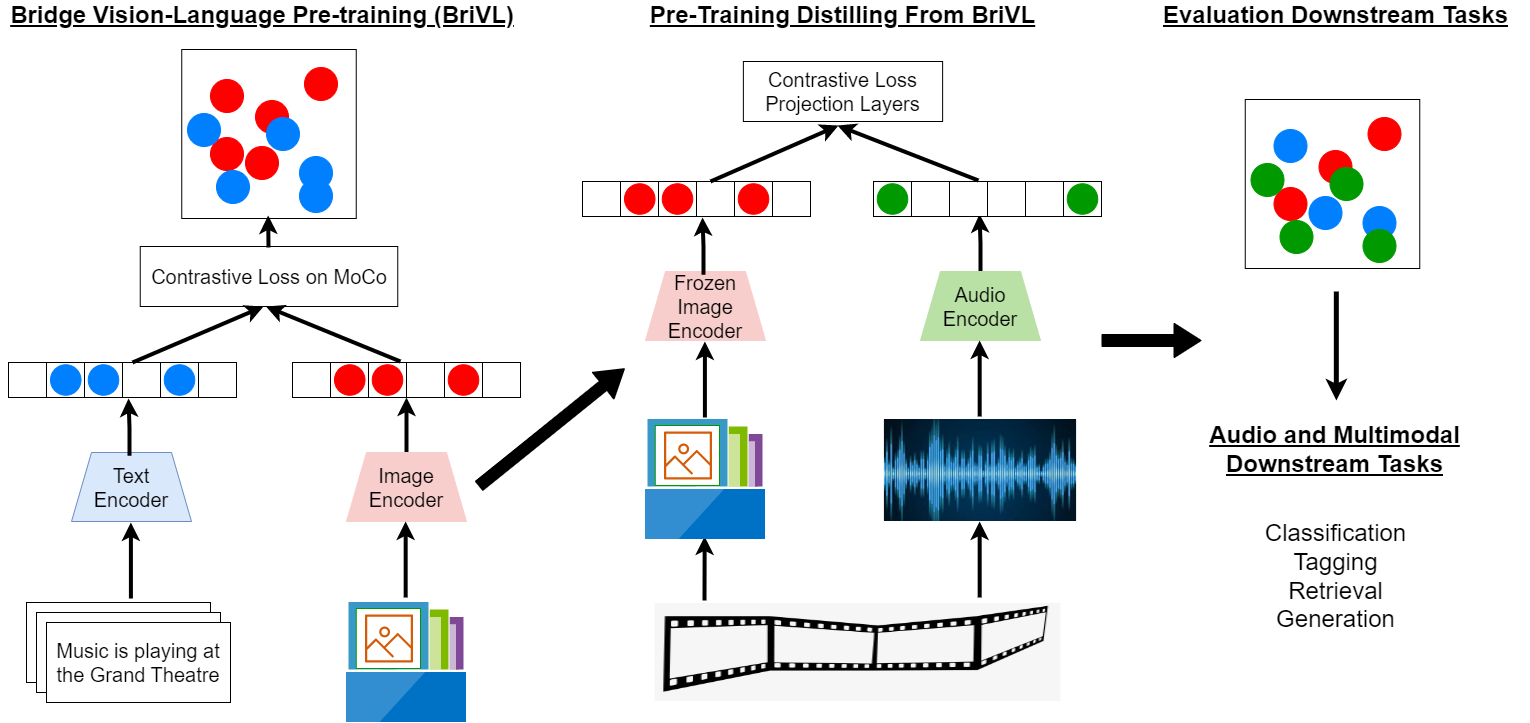}
\caption{Bridge Vision-Language Pre-training (BriVL), and our two-stage approaches including pre-training and evaluation.}
\label{fig:methods}
\end{figure}

\section{Methodology And Experiments}

Our method uses the WavBriVL model to guide the generation of VQGAN \cite{esser2021taming} output images through adversarial neural networks.
This process utilizes meaningful embedding in the embedding space, by calculating the matching score between audio and image to rearrange the image, and this rearrangement idea is like CLIP.
Our code is improved from the official model code and similarity calculation tools\footnote{\url{https://github.com/BAAI-WuDao/BriVL}}.
We found that this method can generate images that are appropriate for a given audio input, as confirmed by feedback from related experiments.
Furthermore, our approach has the advantage of requiring less data compared to other fully supervised models for achieving competitive performance in downstream tasks. 
Specifically, WavBriVL's pre-training is more effective than competitive methods because it does not need to completely re-learn the visual model; 
instead, it only needs to train the audio model. 
This makes the model a promising model for various applications.

Bridging-Vision-and-Language is a model that has been trained on 650 million weak semantic datasets containing both text and images. To achieve this, the researchers created a cross-modal comparison learning algorithm based on MoCo \cite{9157636}, a monomodal comparison learning method. The Memory Bank mechanism was employed to maintain negative sample queues across different training batches, enabling the generation of a large number of negative samples for use in the comparison learning method.
Additionally, BriVL has demonstrated state-of-the-art performance across various multi-modal tasks, including image annotation and zero-shot classification.
As shown in Figure \ref{fig:methods}, we replace the text encoder with the audio encoder by freezing the visual model of BriVL, running the image through it, and training the new model to predict that only the matching image-embedded content is obtained from the audio. 
After the audio encoder is trained, we freeze it and used it for experiments related to WavBriVL image generation to evaluate the results of these experiments.
The assessment tasks include sound volume assessment and qualitative image quality and quantitative assessment.
The rearranged images are all provided by selecting from the 100th epoch of the same 20 text inputs.

\begin{table*}[htb]
\centering
\begin{tabular}{@{}ccl@{}r@{}c@{}}\toprule
Dataset &Task &  Clip (Split) &  Class & Metric \\
\midrule
ESC-50 \cite{piczak2015dataset} &MC/ZS & 2k (5 folds) & 50 & ACC \\
UrbanSound8K \cite{Salamon:UrbanSound:ACMMM:14} &MC/ZS & 8k (10 folds) & 10 & ACC \\
VGGSound \cite{chen2020vggsound} &MC/ZS & 185k & 309 & mAP \\
\midrule
DESED \cite{Turpault2019_DCASE} &AR & 2.5k (valid) & 10 & F1 \\
VGGSound \cite{chen2020vggsound} &CMR & 15k (test) & 309 & MRR \\
\midrule
Clotho \cite{Drossos_2020_icassp} &AC & 5k (evaluation) & & COCO \\
\bottomrule
\end{tabular}
\caption{Downstream tasks, including 1. classification: multi-class (MC), zero-shot (ZS), 2. retrieval: audio (AR) and cross-modal retrieval (CMR), and 3. audio captioning (AC) task, with various of clips, classes, and common metrics.}
\label{tab:audio_tasks}
\vspace{-2.0em}
\end{table*}

\subsection{Dataset for WavBriVL performance test}
\label{ssec:Dataset.test}

Our experiments encompassed a broad range of datasets, incorporating varying numbers of clips and categories. Additionally, we performed diverse tasks such as classification, retrieval, and generation to ensure comprehensive performance evaluation. 
For evaluation, we use relevant metrics detailed in Table \ref{tab:audio_tasks} for each task.
BriVL needs more than 100 A100 graphics cards to train for 10 days, so we don't consider retraining it.
Our training and performance testing are based on the pre-trained model.

\subsection{Dataset for VQGAN}
\label{ssec:Dataset}

In our study, we utilized one video dataset, namely VGG-Sound \cite{chen2020vggsound} to conduct the main experiment. 
The VGG-Sound dataset comprises audio sound clips, comprising 310 video classes and 200,000 audio samples that present challenging acoustic environments and practical noise characteristics. 
These audio samples are associated with non-man-made videos, and our model randomly selected one image from each sample video, with an audio sampling rate of 16kHz. 
We used 20000 video clips for training. Then select another part to evaluate.
It is noteworthy that the original BriVL model used Chinese datasets, while VGG-Sound's datasets are in English. 
Since we only attempted to perform related evaluation tasks based on image generation, the language of the dataset has no significant impact at the moment, which is sufficient for some analysis, and we only consider fine-tuning it.
Moreover, we incidentally cleared the original text-related weights to avoid potential impacts.

\subsection{Feature extraction processing}


Our image and audio encoders utilize EfficientNet-B7 \cite{tan2019efficientnet} as the CNN in the former and WavLM \cite{chen2022wavlm} as the basic transformer in the latter. The self-attention block combines four Transformer encoder layers and an MLP block, incorporating two fully connected layers and one ReLU activation layer.
Our sorting code conducts forward operations on both the audio and image data before generating an image-audio matrix for sorting and output. Additionally, the code provides similarity scores for the processed vector matrix and the indices sorted by score to sort the images based on relevance.

\textbf{Image Encoder.} We adapted BriVL's approach of using random grayscale in the input image with color jitter for data enhancement. We standardized all videos at 1080P resolution and separated images (or 720P if not available), cropping them down to 480 x 480 pixels. To capture patch features and extract them via average pooling, we use a Transformer. To better capture patch feature relationships, we employed a self-attention (SA) block containing multiple Transformer encoder layers, each comprising a multi-head attention (MHA) layer and a feedforward network (FFN) layer \cite{fei2022towards}.
\begin{eqnarray}
&& \mathbf{S}' = \textrm{LayerNorm}(\mathbf{S} + \textrm{MHA}(\mathbf{S})) \\
&& \mathbf{S} = \textrm{LayerNorm}(\mathbf{S}' + \textrm{FFN}(\mathbf{S}'))
\end{eqnarray}
Then, they use the average pooling layer to fuse the extracted patch features:
\begin{equation}
\mathbf{r}^{(i)} = \frac{1}{N_p} \sum_{j=1}^{N_p} \mathbf{S}_j \in \mathbb{R}^{c}
\end{equation}
where $\mathbf{S}_j$ is the $j$-th column of $\mathbf{S}$. A two-layer MLP block with a ReLU activation layer is adopted to project $\mathbf{r}^{(i)}$ to the joint cross-modal embedding space, resulting in the final $d$-dimensional image embedding $\mathbf{z}^{(i)} \in \mathbb{R}^{d}$.

\textbf{Audio Encoder.} In the case of audio input, our method initially converts the original 1D audio waveform into a 2D spectrum that serves as the input for WavLM. To output an embedding, we pool the entire 512-dimensional audio sequence and calculate the WavLM embedding using a weighted average of outputs from all transformer layers.
The WavLM model\footnote{\url{https://github.com/microsoft/unilm/tree/master/wavlm}}, which was inspired by HuBERT, comprises a CNN encoder and a Transformer with $L$ blocks. During training, some frames of the CNN encoder output $\mathbf{x}$ are randomly masked and fed to the Transformer as input. The Transformer is trained to predict the discrete target sequence $\mathbf{z}$, in which each $z_t \in [C]$ represents a $C$-class categorical variable. 
The distribution over classes is parameterized using Equation (1), where $\mathbf{W}^P$ is a projection matrix, $\mathbf{h}^{L}_{t}$ denotes the output hidden state for step $t$, $\mathbf{e}_c$ represents the embedding for class $c$, $\mathrm{sim}(a,b)$ calculates the cosine similarity between $a$ and $b$, and $\tau=0.1$ scales the logit \cite{chen2022wavlm}:
\begin{equation}
    p(c|\mathbf{h}_t)=\frac{\exp(\mathrm{sim}(\mathbf{W}^P\mathbf{h}^{L}_{t},\mathbf{e}_c)/\tau)}{\sum_{c'=1}^{C}\exp(\mathrm{sim}(\mathbf{W}^P\mathbf{h}^{L}_{t},\mathbf{e}_{c'})/\tau)}
\end{equation}
The distribution over classes in extraction method is parameterized using Equation (1), where $\mathbf{W}^P$ represents a projection matrix that is optimized during training. The output hidden state for step $t$, denoted by $\mathbf{h}^{L}_{t}$, is computed by the Transformer with $L$ blocks. The embedding for class $c$ is represented by $\mathbf{e}_c$, and the cosine similarity between $a$ and $b$ is calculated using $\mathrm{sim}(a,b)$. The scaling factor $\tau=0.1$ is used to adjust the logit score.
During fine-tuning, the WavLM embedding as the weighted average of all transformer layer outputs, with the weights learned during this stage. Fine-tuning involves updating the parameters of WavLM to optimize the model for a specific downstream task.



\section{Task 1: WavBriVL Performance Test}
\label{sec:taskhead}

In this chapter, we begin by discussing the training, development, and evaluation process of the WavBriVL model. We use publicly available datasets of varying sizes and tasks, including classification, retrieval, and audio captioning tasks. We compare WavBriVL with some widely used as strong benchmarks in this field, and evaluate its performance in these tasks.
Additionally, we investigate the effect of sound volume on the generated images. We hypothesize that the volume of sounds can influence the generated images. Hence, we explore the influence of sound volume on image features extracted from the sound using the sound correlation model.
We also perform quantitative image analysis to evaluate the performance of WavBriVL compared to previous work, such as S2I and Pedersoli et al. We test model with five categories from VEGAS \cite{vegas} and compare its performance with other methods in terms of generating visually plausible images.
In the experiment, the rearranged images are all provided by selecting from the 100th epoch of the same 20 text inputs.


\subsection{Training, development, and evaluation}
\label{sssec:downstream}

We selected publicly available audio classification data of different sizes, which are generally used for evaluation \cite{cramer2019look}, and also included some audio tasks/data, as shown in table \ref{tab:audio_tasks}, including classification, retrieval and audio captioning. 
ESC-50 \cite{piczak2015dataset} is a simple data set with only 2 thousand samples, while UrbanSound8K \cite{Salamon:UrbanSound:ACMMM:14} is a large environmental data set with 10 categories. 
VGGSound \cite{chen2020vggsound} is a huge set of audio and video materials as we said before.
DESED is used again as an audio extraction (AR) job because DESED can perform sound extraction at the fragment level. Finally, Clotho \cite{Drossos_2020_icassp} is a unique set of audio subtitles.
In addition, in chapter \ref{ssec:quan}, we also carried out other image analysis.


\begin{table*}[htb]
\centering
\scalebox{0.75}{  
\begin{tabular}{@{}c@{\hskip 0.1in}c@{\hskip 0.1in}c@{\hskip 0.1in}c@{\hskip 0.1in}c@{\hskip 0.1in}c@{\hskip 0.1in}c@{\hskip 0.1in}c@{\hskip 0.1in}c@{}}\toprule
& \multicolumn{4}{c}{Classification} & \multicolumn{4}{c}{Retrieval} \\
\cmidrule(lr){2-5} \cmidrule(lr){6-9}
Model & ESC-50 & UrbanSound8K & VGGSound & DESED (AR) & \multicolumn{2}{c}{VGGSound (CMR)} \\
\cmidrule(lr){8-9}
& ACC & ACC & mAP & F1 & A$\rightarrow$I (MRR) & I$\rightarrow$A (MRR) \\
\midrule
Supervise & 0.5200 & 0.6179 & 0.4331 & & \\
OpenL3 & 0.733 & 0.7588 & 0.3487 & 0.1170 & 0.0169 & 0.0162 \\
Wav2CLIP & 0.8595 & 0.8101 & 0.4663 & 0.3955 & 0.0566 & 0.0678 \\
WavBriVL & \textbf{0.9117} & \textbf{0.8832} & \textbf{0.4741} & 0.3720 & \textbf{0.0611} & \textbf{0.0608} \\
\midrule
SOTA & 0.959 & 0.8949 & 0.544 & & \\
\midrule
WavBriVL (ZS) & 0.412 & 0.4024 & 0.1001 & & \\
\bottomrule
\end{tabular}
}
\caption{In the subsequent classification and acquisition work, there will be supervised training, other audio representation modes, OpenL3, and the latest SOTA \cite{guzhov2021audioclip,Kazakos2021SlowFastAuditory}. ZS is based on WavBriVL as a zero sample size model, some of which are derived from the original literature.}
\label{tab:all_tasks}
\vspace{-1.0em}
\end{table*}



For multi-class (MC) classification problems, an MLP-based classifier is employed, with a corresponding number of classes as output.
In DESED, we use the way of simulating WavBriVL and sed\_eval\footnote{\url{https://github.com/TUT-ARG/sed\_eval}} to realize audio retrieval (AR).
At the same time, we also explore the performance of ours when dealing with multimodal tasks, and how to transfer zero samples to other modalities.

\subsection{Comparisons with previous work}

First, we monitor the benchmark by training from scratch on each downlink (with random initialization of the encoder weights).
Next, we compare WavBriVL with other publicly available OpenL3 \cite{cramer2019look} pre-trained on different pretext tasks in OpenL3.
OpenL3 multimodal Self-supervised training with AudioSet.
It serves as a strong benchmark for different audio tasks, such as audio classification and retrieval. We extract features from OpenL3 (512 dim) and WavBriVL (512 dim) and apply the same training scheme to all downstream classification and retrieval tasks.
In the chart, we can see that in the retrieval of classification, we are slightly better than our previous work, with an average increase of about 0.04, and only some deficiencies in AR. 
But it's only about 0.02. We approach or slightly outperform our previous work in retrieval tasks.

In sumary, our model has good effects in both data sets of audio retrieval classification, for the source of our strengths:
In the Classification tasks, on the four datasets, three of us achieved good results close to or exceeding SOTA. 
one of reason may be related to our data, and the other may be the effect of BriVL. 
As for the lack of excellent performance in AR tasks, it may be due to the excessive divergence of the BriVL dataset. 
If we retrain the basic model on a large scale, we may achieve better results.
In the Retrieva tasks, such mrr tasks from A to I, from I to A we have also achieved excellent results, 
which mainly comes from the excellent training effect of the previous two towers model and the pre-training model, the structure of the brief is useful for general with tasks.


\begin{table}[ht]
\begin{minipage}{0.5\linewidth}
\centering
\vspace{1.0em}
\begin{tabular}{llccc}
    \toprule
    &\multirow{2}{*}{Method}&\multicolumn{3}{c}{VEGAS (5 classes)}\\
    \cmidrule{3-5}
    & & R@1 & FID ($\downarrow$) & IS ($\uparrow$)\\
    \cmidrule{1-5}
    (A)& Baseline & 23.10 & 118.68 & 1.19\\
    (B)& S2I & 39.19 & 114.84 & 1.45\\
    (C)& Ours & \textbf{53.55} & \textbf{96.12} & \textbf{3.52}\\
    (D)& SOTA & 77.58 & 34.68 & 4.01\\
    \bottomrule
\end{tabular}
\caption{Compared to the previous three jobs \cite{pedersoli2022estimating,s2i,sungbin2023sound}.}
\label{table:competitive-cvpr}
\end{minipage}\hspace*{0.04\linewidth}
\begin{minipage}{0.45\textwidth}
  \centering
  \includegraphics[width=\textwidth]{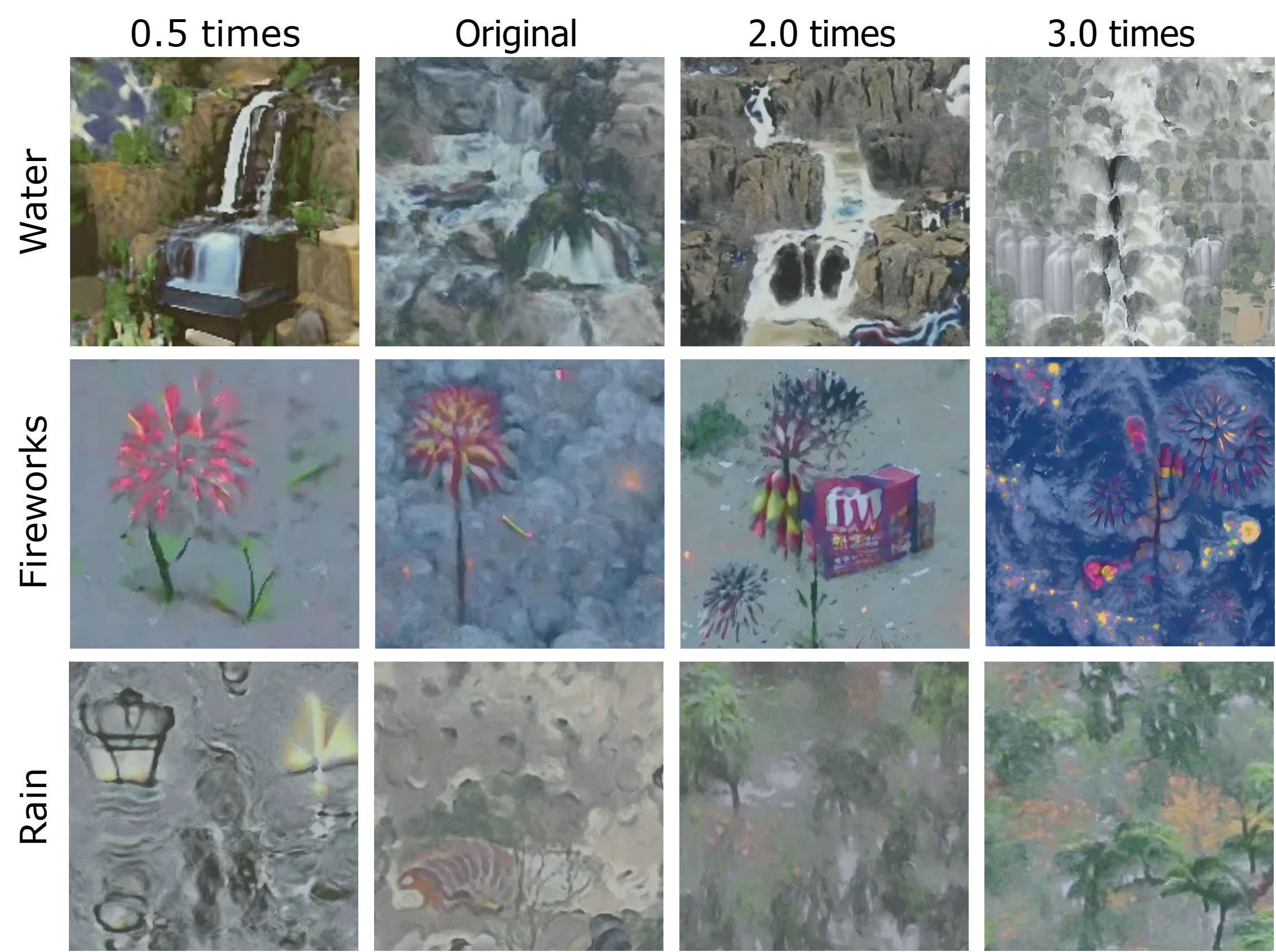}
  \captionof{figure}{The effect of different volume levels on three inputs: water, fireworks, and rain.}
  \label{table:sound}
\end{minipage}
\end{table}

\subsection{Quantitative image analysis}
\label{ssec:quan}
We have carried out a comparative analysis of our proposed model against relevant prior works (S2I\footnote{\url{https://github.com/leofanzeres/s2i}}~\cite{s2i,sungbin2023sound} and \cite{pedersoli2022estimating}). It should be noted that although the S2I was not initially designed for sound-to-image conversion, it leverages a VQVAE-based model for generating sound-to-depth or segmentation. For a fair comparison, we ensured that our model and the one proposed by Pedersoli et al. were trained using the same training set up as S2I, including five categories in VEGAS. As reported in Table \ref{table:competitive-cvpr}, our proposed model outperforms most of the other models while generating visually enthralling and recognizable images, thanks to the amalgamation of visually enriched audio embeddings and a potent image generator.
Our method outperforms the others which using GAN both qualitatively and quantitatively in the VEGAS dataset.

\section{Task 2: Speech Generation Picture}
\label{sec:majhead}

In this chapter, our objective was to enhance the detectability of the shared embedded space of WavBriVL through visual analysis. 
To accomplish this, we employed VQGAN \cite{esser2021taming} to generate images with audio guidance.
To generate these images, we matched them against the input audio and assessed whether BriVL "approved" of the output. 
In the case of a mismatch, feedback was provided to VQGAN\footnote{\url{https://github.com/CompVis/taming-transformers}} to refine subsequent image generation attempts. 
Both VQGAN and WavBriVL were kept frozen during the performance testing process, as is standard practice.


\begin{figure}[ht]
\centering
\includegraphics[width=\linewidth]{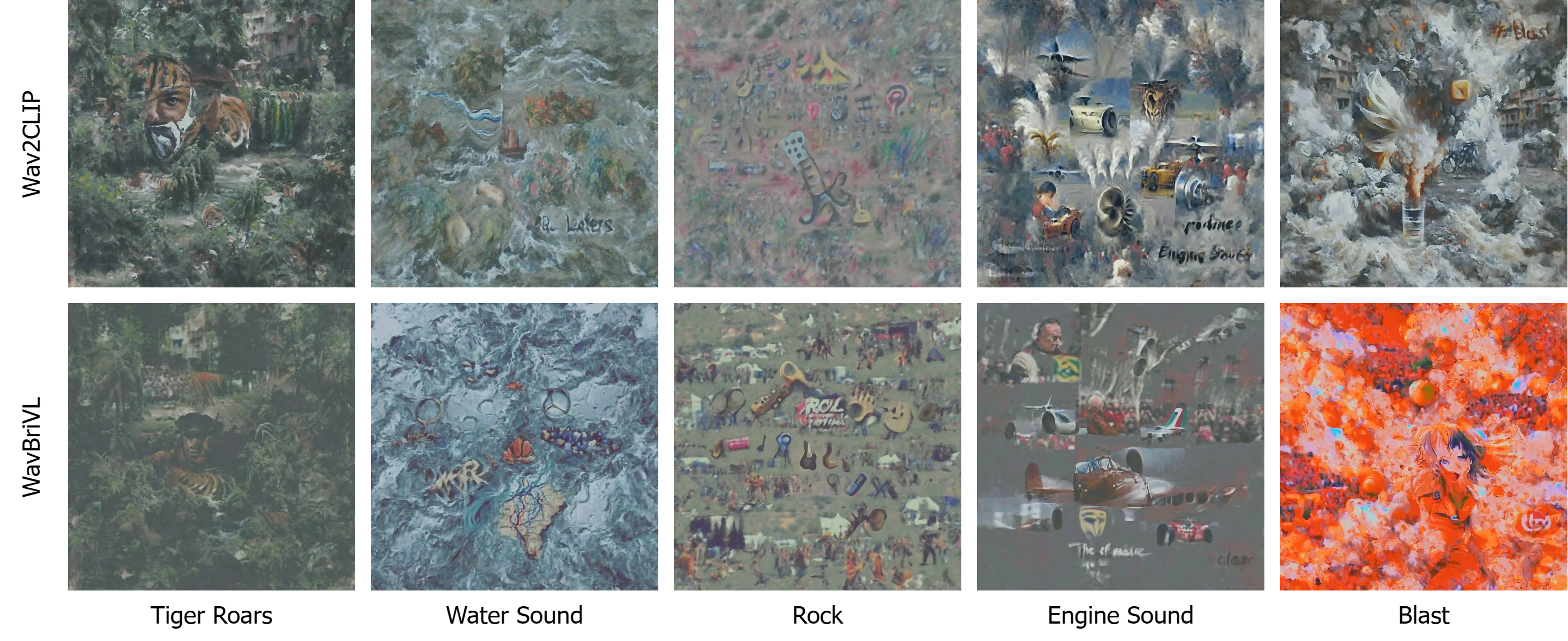}
\caption{Images generated from five audio in VGG-Sound \cite{chen2020vggsound}. Top: Wav2CLIP, Bottom: WavBriVL - The x-axis is the audio based on which the rearrangement is based, and the rearranged images are generated using the same relevant text.}
\label{fig:compare2}
\vspace{-1.0em}
\end{figure}

\subsection{Sound volume}
As shown in Figure \ref{table:sound}, in order to ascertain the robustness of our approach in acquiring an understanding of the correlation between auditory stimuli and visual representations, we conducted a study investigating the impact of varying sound volumes on the quality of the generated images. 
To accomplish this, we manipulated the sound volume levels during the test phase and extracted features from the corresponding sound files. 
Subsequently, these modified sound features were employed as input into our model and improved the similarity calculation tool, which had been trained on a standardized volume scale.
The final three sets of images can prove our hypothesis that the magnitude of different volume levels is usually positively correlated with the effects and meanings displayed in the images.

\subsection{Comparison with previous work}
\label{ssec:cwpw}

In previous work, Wav2CLIP also tried to generate images by audio.  
Figure \ref{fig:compare2} shows two group of pictures generated by Wav2CLIP and WavBriVL using English audio. We can see that the style of the text generation diagram is similar, which is because they all use the same GAN generation model. 
However, in detail, they have their own characteristics: for example, in their understanding of "Blast", the explosion in Chinese videos is often accompanied by an overwhelming number of Chinese New Year videos. 
Therefore, it can be seen that when the image with high similarity of "audio-image matrix" is finally selected by our model, the preference is to represent the festive orange, red firecracker skin, and other elements. 
This is not the difference caused by translation, because the input is the same file, which can show that BriVL has more self-characteristics, and the other pictures have similar characteristics;
In underwater sound input, we can see that their generated images are similar. 
After all, natural underwater sound is more common but more coherent. 
This reflects the argument/advantage of "BriVL generated images are more integrated than CLIP just stacking elements" mentioned in BriVL's original paper.
\begin{table}[ht]
\begin{minipage}{0.5\textwidth}
\centering
\vspace{1.0em}
\begin{tabular}{llll}
\hline
Options        & Positive & Negative & Neither \\ \hline
AudioCLIP & 78\%  & 14\%  & 8\%   \\ \hline
WavBriVL & 81\%  & 15\%  & 4\%   \\ \hline
\end{tabular}
\caption{Human scores on correlation between sounds and images, both AudioCLIP and Wav2BriVL use GAN generated images.}
\label{table:human_evaluation_correlation}
\end{minipage}\hspace*{0.04\linewidth}
\begin{minipage}{0.45\textwidth}
  \centering
  \includegraphics[width=\textwidth]{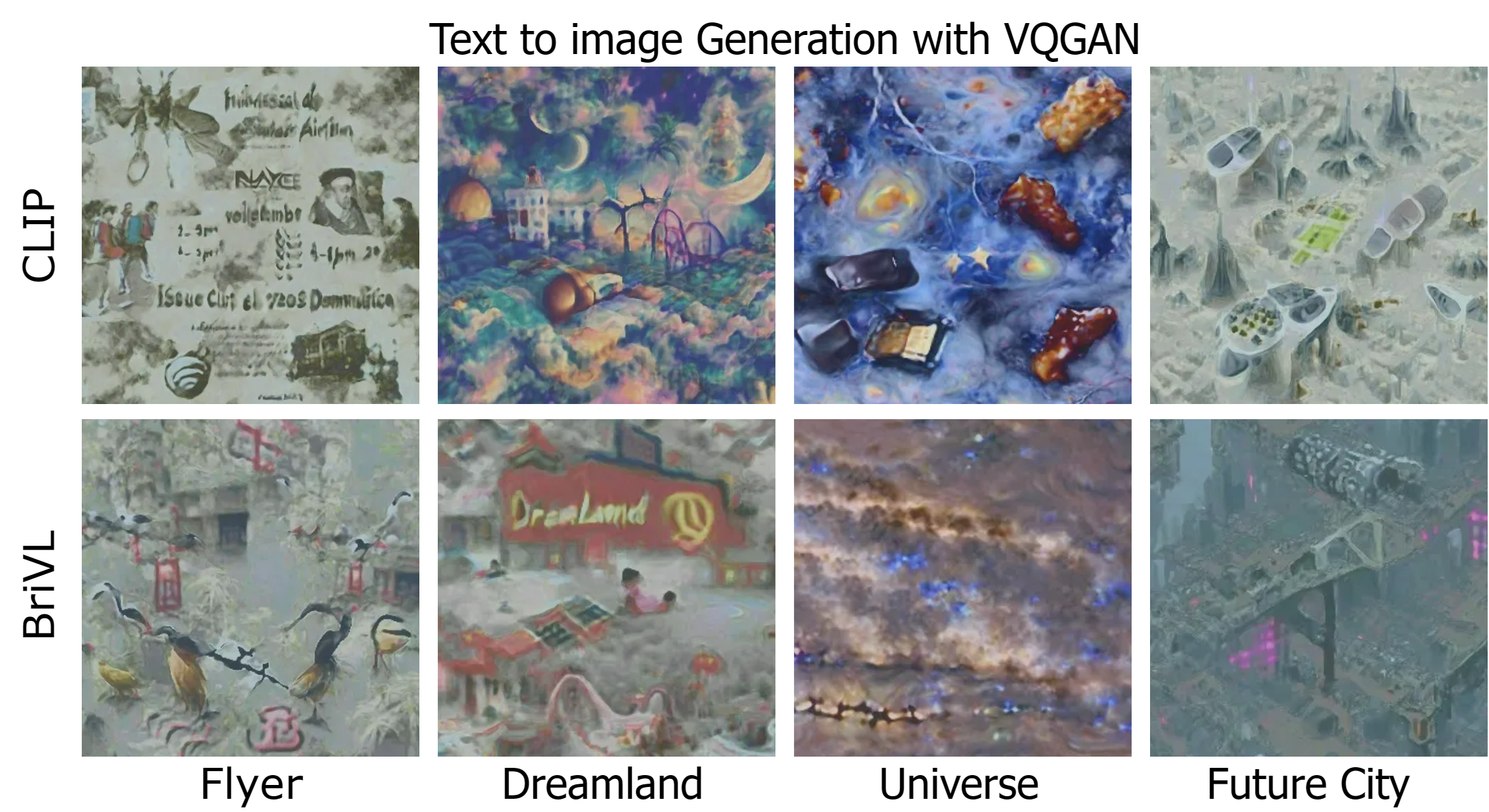}
  \captionof{figure}{Examples of CLIP (top) and BriVL (bottom) to image generation from text, BriVL's labels in x-axis are translated.}
  \label{fig:compare1}
\end{minipage}
\vspace{-1.0em}
\end{table}


The audio generation process exhibits two distinct characteristics in the two models. 
The first characteristic is convergence, as evident from the similarity of images produced by both models. 
This can be attributed to the training set being composed of English images, which dampens the stylistic differences between the two models. 
The second characteristic is divergence
At the same time, in the process of audio guidance and selection, the final result also shows the characteristics of divergence.
As depicted in Figure \ref{fig:compare2}, where the images are more imaginative compared to those in Figure \ref{fig:compare1}. 
This divergence can be explained by two factors: firstly, BriVL's weak semantic text image dataset has a higher imaginative capacity, and secondly, the audio itself has a strong ability to diverge, thereby enhancing the model's associative capabilities.

\subsection{Correlation between sounds and images}
In Figure \ref{fig:compare2}, We show that we can generate better images and interpretable correlations.
However, it only demonstrate their authenticity, but we wanted to assess the relationship between sounds and images. 
To accomplish this, we conducted a test like previous work~\cite{ilharco-etal-2019-large,8682383} where participants were asked to select the image that was most closely linked to a given sound from a set of two images. 
Each of these images was conditioned on a different sound class, so if our model could generate images that were relevant to a particular sound class, then participants would select the corresponding image that was generated from the inputted sound rather than an image generated from a different class of sampled sounds.

Table \ref{table:human_evaluation_correlation} displays the findings of the study, with the options representing the participants' choices. 
A positive option indicates that participants selected the image generated from the sound they heard, while a negative option suggests that they chose the image generated from a different class. 
The "neither" option indicates that participants believed that neither image could represent the sound they heard. 
The results (repeat three times to take the average integer) reveal that the majority of participants believed that the images generated by our model were correlated with the input sounds, demonstrating our model's capability to produce images that are related to the given sounds.


\section{Summary \& Conclusions}
\label{sec:refs}

In this manuscript, we present a method that employs BriVL to extract audio representations for the purpose of audio-guided GAN image generation. 
Our proposed method has been shown to produce suitable audio representations, as demonstrated by its reliability in sound volume assessment, and quantitative and qualitative image quality assessment. 
In the future, we plan to study a model suitable for more modalities, develop the model in greater depth, explore its potential in more tasks, and compare it with other advanced models.
In addition, we will consider trying other Generative model and further using audio representation (such as directly generating images instead of indirectly generating them now) as the work of the next version.
\bibliographystyle{splncs04}
\bibliography{anthology,custom,others}

\end{document}